# Real-time wavefront-shaping through scattering media by all optical feedback

**Micha Nixon[†], Ori Katz[†], Eran Small, Yaron Bromberg, Asher A. Friesem,**

**Yaron Silberberg, and Nir Davidson[*].**

*Department of Physics of Complex Systems, Weizmann Institute of Science, Rehovot 76100, Israel.*

[†] These authors contributed equally to this work

*nir.davidson@weizmann.ac.il

**Focusing light through dynamically varying heterogeneous media is a sought-after goal with important applications ranging from free-space communication to nano-surgery. The underlying challenge is to control the optical wavefront with a large number of degrees-of-freedom (DOF) at timescales shorter than the medium dynamics. Recently, many advancements have been reported[1-15], following the demonstration of focusing through turbid samples by wavefront-shaping, using spatial light modulators (SLMs) having >1000 DOF[2]. Unfortunately, SLM-based wavefront-shaping requires feedback from a detector/camera and is limited to slowly-varying samples[13]. Here, we demonstrate a novel approach for wavefront-shaping using all-optical feedback. We show that the complex wavefront required to focus through highly scattering samples, including thin biological tissues, can be generated at sub-microsecond timescales by the process of field self-organization inside a multimode laser cavity, without requiring electronic feedback or SLMs. This wavefront-shaping mechanism is more than $10^5$ faster than state-of-the-art[13], reaching the timescales required in many applications.**

The ability to focus light through complex, inhomogeneous media has been extensively investigated during the past decades, mainly in order to overcome the deleterious effects of atmospheric turbulence in applications such as astronomical observations, LIDAR, and free-space optical communications[16]. Recently, exciting developments in high-resolution wavefront-shaping using computer-controlled SLMs have been reported[1-15]. These have enabled high contrast focusing even through turbid, nearly-opaque samples, where the light is scattered to complex speckle patterns with a number of scattered modes greatly exceeding the number of controlled DOF[2, 17]. Following the pioneering demonstration of spatial focusing by Vellekoop et al.[2], wavefront-shaping has been exploited also for surpassing the diffraction limit in scattering media[3], for improving imaging[9-11, 15], and for controlling the scattered light in both space and time[6, 7] as well as manipulating its polarization properties[12]. Unfortunately, these developments all rely on computer controlled SLMs and electronic feedback, and so are fundamentally limited by relatively slow response times that range from tens of milliseconds to hundreds of microseconds, to control a single DOF[2, 13]. Consequently, to-date, high resolution wavefront-shaping has not been proven useful for focusing through samples that evolve on timescales shorter than a fraction of a second, such as the ones required for focusing inside live biological tissues or liquid suspensions.

Here we present a novel all-optical approach to high-resolution wavefront-shaping that is capable of focusing light through rapidly varying inhomogeneous media at sub-microsecond timescales, without requiring computer-controlled SLMs or electronic feedback. Our approach relies on the self organization of the optical field inside a multimode laser cavity to generate the optimal wavefront that forms a sharp focus through a scattering medium. This is achieved by utilizing a reflection from a small retro-reflecting target that is placed behind the scattering sample as a coherent all-optical feedback (Fig.1a). This coherent feedback is employed to initiate lasing through the medium in a lasing-state that focuses the maximum light power on the target. The reason for the formation of the tight focus on the target is that the lasing process naturally selects the lasing state with the minimal lasing threshold, i.e. the state with minimal loss. In our cavity design (Fig.1a) this lasing state is the optical field with a complex wavefront that compensates for the scattering of the medium, and that is focused on the target with minimal loss. In the presence of the target's feedback all other lasing modes are suppressed via the process of mode-competition in the laser cavity. The physics behind this focusing mechanism is intimately related to lasing in random media[18-20], and in particular to the role of mode-competition and coherent feedback in random lasers[18]. Specifically, in our experimental implementation (Fig.1a) the coherent feedback to the lasing state with tight focusing dominates the competition over the incoherent feedback from other scattered modes[19]. We note that a coherent reflective feedback is used for focusing through inhomogeneous media in iterative time-reversal techniques in acoustics[21, 22] and self phasing antenna arrays[23], but there focusing is achieved through phase-conjugation, a completely different physical mechanism.

To experimentally investigate, in a controllable manner, the potential of our approach for focusing through highly scattering media, we used a unique self-imaging *('degenerate')* laser cavity design[24] (Fig. 1a), which supports many transverse lasing modes and where the coupling between the modes can be readily controlled and manipulated[25] (see Methods). The cavity was comprised of a flash-lamp pumped Nd-YAG gain medium, two flat cavity mirrors, and two lenses in a 4*f* telescope arrangement (f=400mm). The 4*f* telescope assures that any field distribution will be reimaged on itself after each cavity roundtrip; hence any field is an eigenmode of the degenerate cavity. A controllable aperture placed at the focal plane of the telescope and centre of the cavity served as the target for light focusing. The target's effective reflective feedback and loss to unfocused modes was controlled by varying the aperture diameter (see Methods). The intensity pattern at the target plane was detected by sampling a small portion of the light impinging on the target with a CCD camera, located outside the cavity. To focus through a scattering sample, we placed the sample inside the laser cavity, close to one of the cavity mirrors. The naturally occurring phenomena of mode-competition selects the lasing state with minimal losses, i.e. the appropriate amplitude and phase distribution which passes through the scattering sample but is also tightly focused through the pinhole aperture. In this manner, focusing is achieved by the laser itself and occurs on sub-microseconds lasing timescales, without any further manipulations.

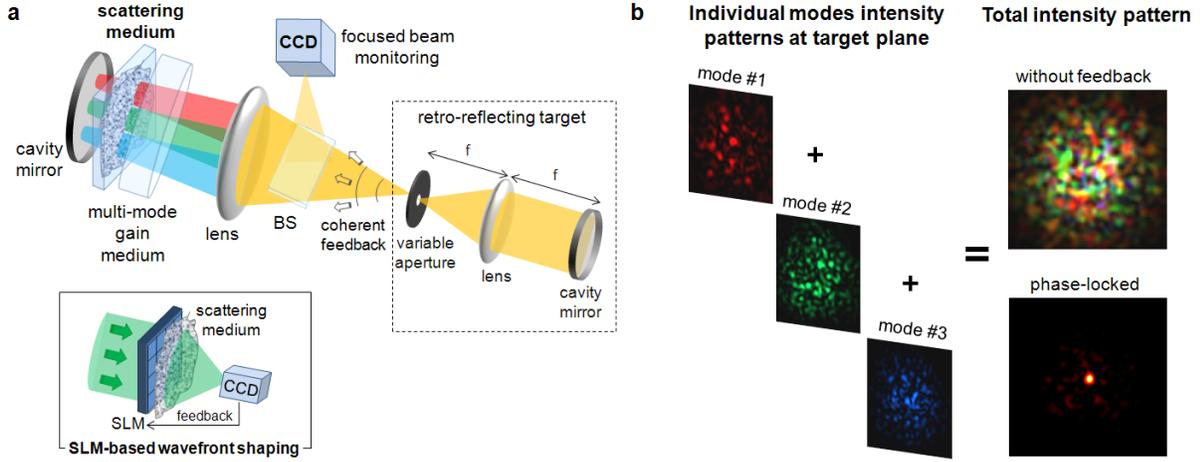

**Figure 1. Wavefront shaping through scattering media by all-optical feedback**. **a,** Experimental setup: The setup is composed of a degenerate *self-imaging* cavity that is based on a 4-f telescope configuration, supporting thousands of independent spatial lasing modes (depicted schematically by red, green and blue beams). A scattering medium is placed inside the laser cavity and a controlled aperture placed at the cavity's centre serves as the target for light focusing, by ensuring that only the focused in-phase component of the scattered light hitting the target are fed-back to the gain medium at each cavity round-trip. A small amount of light is directed towards a CCD camera by a beam sampler (BS) for monitoring the intensity pattern at the target plane. In analogy to the conventional SLM-based wavefront-shaping approach (inset), where focusing is achieved by controlling the spatial phases of the SLM pixels, the focusing in our system is achieved by 'locking' the relative phases and *frequencies* of the independent lasing modes. **b,** Illustration of the focus formation: without the feedback (i.e. with no aperture), each scattered laser mode operates at an arbitrary phase, amplitude and frequency (left insets 1-3) and the total intensity pattern at the target plane is a random one (top-right inset). However, when the laser modes are coupled by the reflective feedback, they lock onto the same frequency and correct phases to create an intensity peak on the target (bottom-right inset).

As a first experiment, we placed an optical diffuser adjacent to the back cavity mirror. Simple focusing of a reference plane-wave from an external CW laser source through the diffuser resulted in a random speckle pattern at the target plane, with no appreciable unscattered, zero-order component (Fig.2a). In contrast, the laser operates in a lasing state that forms a significantly enhanced intensity peak through the diffuser, refocusing the scattered light on the target (Fig. 2b). Interestingly, the lasing and the refocused intensity peak were stable and insensitive to transverse shifts of the diffuser or to tilts of the adjacent cavity mirror within the diffuser scattering angle, whereas without the diffuser, the cavity required precise careful alignment to achieve optimal lasing. Quantitatively, comparing the intensity cross-sections of the lasing-pattern to that of the reference plane-wave focusing gives a relative intensity enhancement factor of $\eta \approx 80$ on the target area (Fig.2c). In conventional wavefront-shaping, this enhancement $\eta$ is given by the number of controlled DOF, $N_{DOF}$, over the number of intensity enhanced speckles $N_{enh}$, i.e. $\eta \approx N_{DOF}/N_{enh}$ [1,2]. In analogy to conventional wavefront-shaping, the measured enhancement factor and the fact that the number of intensity-enhanced speckles on our target is $N_{enh} \approx 15$ suggests the control of over 1,000 DOF in our experiment. This result is in accordance with the number of supported transverse modes in our cavity (see Supplementary Information). In addition, we found that as in conventional wavefront-shaping through turbid media[1,2] even though the intensity enhancement is large, only a small fraction of the total energy is focused on the target. In the results presented in Fig.2, the laser refocuses approximately 10% of the energy on the target, which has an area approximately $10^3$ times smaller than the initial spread of the beam. As expected, we observed an increase in the enhancement as $N_{enh}$ is lowered, e.g. by reducing the aperture diameter. However, the fraction of focused energy also decreases, increasing lasing threshold, and suggesting that the size of the aperture affects $N_{DOF}$.

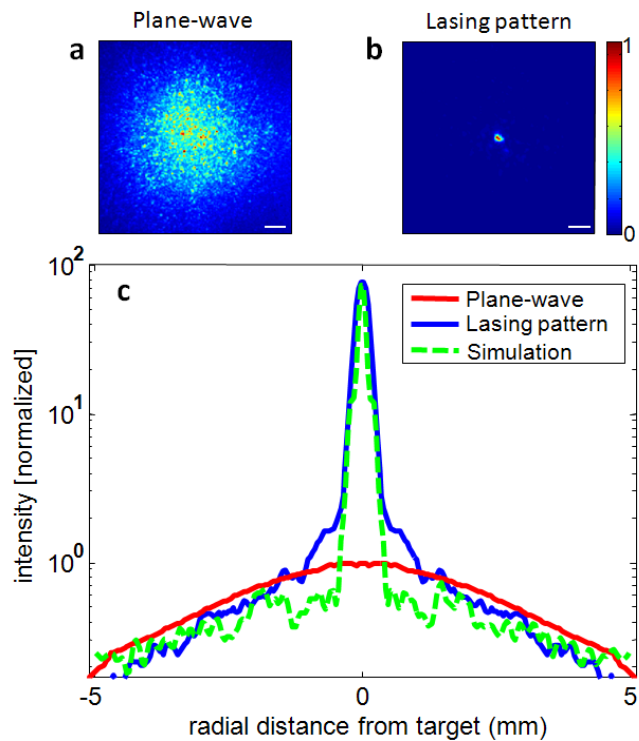

**Figure 2. Focusing through an optical diffuser with all-optical feedback, experimental results. a,** Intensity distribution at the target plane obtained by conventional focusing of an incident plane-wave through the diffuser. **b,** Intensity distribution at the target plane of the lasing pattern obtained by all-optical feedback, demonstrating focusing on the 0.6 mm target through the diffuser. **c,** Comparison between the intensity cross-sections of the results shown in (a,b). The plane-wave focusing is shown in red, and the lasing profile is shown in blue (cross-sections are taken by tangentially integration the images that are normalized to have the same total power); Dashed green curve, numerical simulation results for the lasing profile through the diffuser (see text), Scale bars are 1mm.

To understand the physics behind the mechanism by which the all-optical feedback is used to focus light in our experiments, it is constructive to draw a comparison with the conventional approach to wavefront shaping using SLMs[2] (Fig. 1a, inset). In the latter, the signal from a small detector placed behind a scattering medium (e.g. from a single pixel of a camera) is used as the feedback signal for an iterative optimization algorithm, aiming at maximizing the detected intensity by manipulating the spatial phases of a monochromatic laser beam. In an analogy, the complex-patterned laser beam impinging on the scattering medium in our scheme, can be decomposed to an array of localized laser modes, each with its own independent phase, which effectively act as independent lasers[25]. Following this analogy, the array of independent laser modes mimics the array of controllable pixels used in conventional SLM-based wavefront-shaping. This analogy is of course not perfect as it neglects the fact that the random spatial phase modulation induced by the scattering medium also alter, locally, the cavity`s optical length giving rise to frequency shifts of the different laser modes. As a consequence, focusing the light through the random medium inside the cavity is achieved by simultaneously locking of both the relative phases *and frequencies* of the array of laser modes (Fig.1b). In this laser-array analogy, the focus is formed by a state of mutual coherence among the laser modes which arises as result of coupling induced by diffraction from the small reflecting target. Specifically, the diffraction causes light to couple from one laser mode to another thereby phase-locking them through the process of frequency pulling[26] (see Supplementary information). The lasing state with the maximal reflectivity from the target will be selected via mode-competition, thereby focusing the optical power on the target, which then effectively serves as the front cavity-mirror of the lasers array.

To investigate the speed at which our wavefront shaping approach can respond to dynamically varying turbidity, we replaced the static diffuser with a diffuser that was mounted on the edge of a rapidly rotating wheel. By monitoring the lasing intensity distribution at the target plane for different rotating speeds, we could evaluate the focusing temporal dynamics. Theoretically, focusing is expected to occur on timescales compatible with the phase-locking time of the coupled lasers, which is governed by the photon cavity lifetime[27, 28], and is tens of nanoseconds in our system. The experimental results of this investigation are presented in Fig.3. It shows the normalized peak intensity on the target as a function of the linear velocity of the diffuser, $v$, and correspondingly the phase decorrelation time for the composite cavity mode: $\tau_{corr} \approx \sigma_{corr}/v$, where $\sigma_{corr} \sim 50\mu m$ denotes the diffuser's spatial correlation length, which is larger than the individual lasing-mode diameter. As evident, the focused intensity-peak is essentially insensitive to the motion of the diffuser, to within 20%, and tight focusing occurs even for the maximum rotation speed allowed by our experimental system $v>80m/s$, which corresponds to $\tau_{corr} \approx \sigma_{corr}/v < 620ns$. Such a timescale is not just considerably faster than the typical dynamics of biological samples, but is also faster than the typical pixel dwell-time in most imaging applications, including laser-scanning microscopy and OCT, giving rise to potential applications which were hitherto not feasible.

To numerically study our approach, we used the "lasers-array" physical picture described in the previous paragraphs to develop a simplified model, which helped in obtaining further insight into the focusing mechanism in our experiment. In this model, we spanned the laser-modes in a basis of small circular modes arranged on a tightly packed triangular lattice covering the surface of the gain aperture (Supplementary Fig.1c). The thin scattering diffuser is modelled by a random phase-mask with a spatial correlation-length matching the diffuser scattering angle. We then calculated, for this random phase-mask, the lasing steady-state solution of the entire array, taking into account the coupling between modes as well as mode-competition and gain saturation. This step was accomplished by propagating the modes repeatedly through the simulated cavity, using similar parameters as in the experiment (for more details see Supplementary materials). The numerical results for the steady-state lasing intensity cross-section at the target-plane are presented in Fig.2c by a dashed green curve. As evident, the numerical results are in very good agreement with the corresponding experimental results. The numerical simulation results show that focusing is indeed achieved by complex shaping in both spatial amplitude and phase (see Supplementary Material).

Finally, we applied our technique to focus light through a thin scattering biological sample. Specifically, we replaced the diffuser with a slice of approximately $200\mu m$ thick chicken breast in water and Glycerol solution, placed between two microscopes slides. As with the optical diffuser, the light of an incident focused plane-wave was scattered to a random speckle pattern without a noticeable ballistic component (Fig .4a). However, the lasing intensity pattern maintained an effective tight focus through the scattering tissue on the target (Fig. 4b).

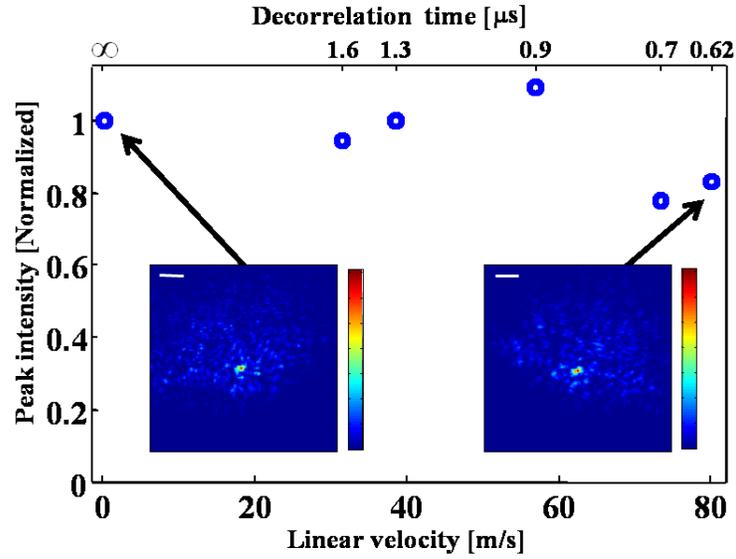

**Figure 3. Focusing through a time-varying diffuser.** Comparison of the focused peak intensity through a rapidly rotating diffuser to that of the stationary diffuser. As focusing is achieved at sub microsecond timescales (the photon cavity lifetime), no appreciable degradation of the focus intensity is apparent even at very fast decorrelation times. Insets: intensity patterns measured for the static case (left) and for a rapidly rotating diffuser with a decorrelation time of 620 ns (right); Scale bars are 1mm.

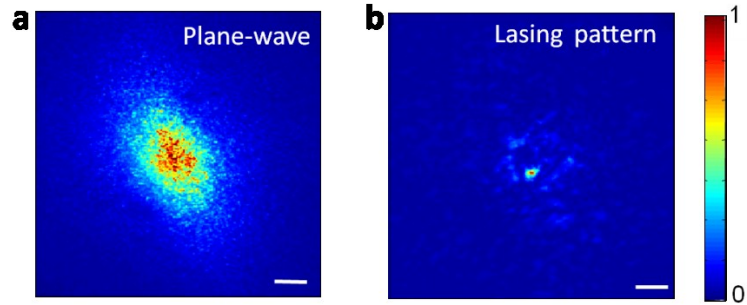

**Figure 4. Focusing through a ~200μm thick chicken-breast sample. a,** Intensity distribution at the target plane for conventional focusing of a plane-wave from an external source through a ~200μm thick slice of chicken breast. **b,** Intensity distribution at the target plane of the lasing pattern, demonstrating sharp focusing of the light through the thin slice of chicken breast on the target; Scale bars are 1mm.

In conclusion, we have demonstrated an all-optical technique for wavefront-shaping, focusing light through highly scattering media at unprecedented speeds, without requiring the use of adaptive algorithms, SLMs[2], or electronic feedback[16]. In addition to its unmatched speed, the all-optical technique has the potential to drastically increase the number of controlled DOFs (number of transverse modes) without trading-off speed as in other optimization-based techniques[2]. Attractive applications exist in free-space optical communication through atmospheric turbulence[29], where the two halves of the cavity could be divided between the transmitter and the receiver. Biomedical applications such as focused deep-tissue laser therapy are especially appealing but would require further research for the appropriate optical setup. An exciting path in this field may lie in combining the use of biological tissue as the gain medium itself[30]. Finally, it would be interesting to study and exploit the complex temporal dynamics of this system, e.g. by combining a saturable absorber at the target it may be possible to achieve temporal focusing ("mode locking"), in addition to the demonstrated spatial focusing.

**Methods**

**Experimental.** All results of focusing through scattering samples were achieved by placing the samples inside of the laser cavity presented schematically in Fig. 1(**a**), next to the left cavity mirror. The cavity was comprised of a 10 *mm* diameter Nd-Yag crystal gain medium that can support thousands of independent transverse modes, and high reflectively flat cavity mirrors placed at either ends of the laser cavity. Two lenses with *f=400mm* focal lengths were placed inside the laser cavity in a 4*f* telescope arrangement which images the back mirror plane onto the front mirror plane, thereby ensuring that any transverse mode is an eigenmode of the cavity[24]. A pinhole aperture placed at the focal plane in-between the lenses served as the target for light focusing. Pinhole sizes varied between 0.4mm to 0.6mm. The diffuser used in the experiment presented in Fig.2 was a 1° Newport light shaping diffuser. The optical diffuser used in Fig. 4 was a thin diffusive plastic sheet with ~1° scattering angle and no zero-order ballistic component. The plane-wave illumination which was used as a reference for comparison to conventional focusing was generated by a single-mode Yb cw fiber laser lasing at 1.064µm.

**References**

1. Mosk, A.P., Lagendijk, A., Lerosey, G. & Fink, M. Controlling waves in space and time for imaging and focusing in complex media. *Nat Photon* **6**, 283-292 (2012).
2. Vellekoop, I.M. & Mosk, A.P. Focusing coherent light through opaque strongly scattering media. *Opt Lett* **32**, 2309-2311 (2007).
3. Vellekoop, I.M., Lagendijk, A. & Mosk, A.P. Exploiting disorder for perfect focusing. *Nat Photon* **4**, 320-322 (2010).
4. Popoff, S.M. et al. Measuring the transmission matrix in optics: an approach to the study and control of light propagation in disordered media. *Phys Rev Lett* **104**, 100601 (2010).
5. Cizmar, T., Mazilu, M. & Dholakia, K. In situ wavefront correction and its application to micromanipulation. *Nat Photon* **4**, 388-394 (2010).
6. Katz, O., Small, E., Bromberg, Y. & Silberberg, Y. Focusing and compression of ultrashort pulses through scattering media. *Nat Photon* **5**, 372-377 (2011).
7. Aulbach, J., Gjonaj, B., Johnson, P.M., Mosk, A.P. & Lagendijk, A. Control of Light Transmission through Opaque Scattering Media in Space and Time. *Physical Review Letters* **106**, 103901 (2011).
8. McCabe, D.J. et al. Spatio-temporal focusing of an ultrafast pulse through a multiply scattering medium. *Nat Commun* **2**, 447 (2011).
9. Katz, O., Small, E. & Silberberg, Y. Looking around corners and through thin turbid layers in real time with scattered incoherent light. *Nat Photon* **6**, 549-553 (2012).
10. Wang, Y.M., Judkewitz, B., DiMarzio, C.A. & Yang, C. Deep-tissue focal fluorescence imaging with digitally time-reversed ultrasound-encoded light. *Nat Commun* **3**, 928 (2012).
11. Si, K., Fiolka, R. & Cui, M. Fluorescence imaging beyond the ballistic regime by ultrasound-pulse-guided digital phase conjugation. *Nat Photon* **6**, 657-661 (2012).
12. Guan, Y., Katz, O., Small, E., Zhou, J. & Silberberg, Y. Polarization control of multiply scattered light through random media by wavefront shaping. *Opt. Lett.* **37**, 4663-4665 (2012).
13. Conkey, D.B., Caravaca-Aguirre, A.M. & Piestun, R. High-speed scattering medium characterization with application to focusing light through turbid media. *Opt. Express* **20**, 1733-1740 (2012).
14. Hsieh, C.-L., Pu, Y., Grange, R. & Psaltis, D. Digital phase conjugation of second harmonic radiation emitted by nanoparticles in turbid media. *Opt. Express* **18**, 12283-12290 (2010).
15. Popoff, S., Lerosey, G., Fink, M., Boccara, A.C. & Gigan, S. Image transmission through an opaque material. *Nat Commun* **1**, doi:10 1038/ncomms1078 (2010).
16. Tyson, R.K. Principles of adaptive optics, Edn. 3rd. (Academic Press, Boston; 2010).
17. Sebbah, P. Waves and imaging through complex media. (Kluwer Academic Publishers, Dordrecht ; Boston; 2001).
18. Cao, H. Review on latest developments in random lasers with coherent feedback. *Journal of Physics A: Mathematical and General* **38**, 10497 (2005).
19. Cao, H. Lasing in random media. *Waves in Random Media* **13**, R1-R39 (2003).
20. Wiersma, D.S. The physics and applications of random lasers. *Nat Phys* **4**, 359-367 (2008).
21. Montaldo, G., Tanter, M. & Fink, M. Real time inverse filter focusing through iterative time reversal. *J Acoust Soc Am* **115**, 768-775 (2004).
22. Fink, M. Time Reversed Acoustics. *Physics Today* **50**, 34-40 (1997).
23. Skolnik, M. Self-phasing array antennas. *IEEE Transactions on Antennas and Propagation* **12**, 142-149 (1964).
24. Arnaud, J. Degenerate optical cavities. *Applied optics* **8**, 189-195 (1969).
25. Nixon, M. et al. Synchronized cluster formation in coupled laser networks. *Physical Review Letters* **106**, 223901-223901 (2011).

26. Fabiny, L., Colet, P., Roy, R. & Lenstra, D. Coherence and phase dynamics of spatially coupled solid-state lasers. *Physical Review A;(United States)* **47** (1993).
27. Kanter, I. et al. Synchronization of Mutually Coupled Chaotic Lasers in the Presence of a Shutter. *Physical Review Letters* **98**, 154101 (2007).
28. Xu, J., Li, S., Lee, K.K. & Chen, Y.C. Phase locking in a two-element laser array: a test of the coupled-oscillator model. *Opt. Lett.* **18**, 513-515 (1993).
29. Xiaoming, Z. & Kahn, J.M. Performance bounds for coded free-space optical communications through atmospheric turbulence channels *IEEE Transactions on Communications* **51**, 1233-1239 (2003).
30. Gather, M.C. & Yun, S.H. Single-cell biological lasers. *Nat Photon* **5**, 406-410 (2011).

## Acknowledgements

This work was supported by the Israel Science Foundation, ERC advanced grant QUAMI, and the Crown Photonics Center.

## Author contribution

All authors contributed to designing the experiments and writing the manuscript. M.N., O.K. and E.S performed the experiments. M.N. and O.K. analyzed the results and performed numerical simulations.

## Additional information

The authors declare no competing financial interests.

Supplementary Information for:

# Real-time wavefront-shaping through scattering media by all optical feedback

M. Nixon[†], O. Katz[†], E. Small, Y. Bromberg, A. A. Friesem, Y. Silberberg and N. Davidson.

*Dept. of Physics of Complex Systems, Weizmann Institute of Science, Rehovot 76100, Israel.*

In this section we describe the numerical model used to simulate the experimental system, and which its results are presented in Fig.2c of the main text. This model was developed in order to study and verify the mechanism by which the all-optical feedback is used to generate the optimal wavefront that focuses through the scattering samples.

As discussed in the manuscript, the fact that a diffuser is set inside the laser cavity amounts to having a phase-mask with different optical paths at the different portions of the beam. This spatially modulates the effective optical cavity length at different spatial positions and therefore the lasing frequencies at the different spatial portions of the beam (each corresponding to the mean local phase in that region of the diffuser). A simple route to numerically calculate the steady-state lasing solution under these conditions and to verify that indeed multimode lasing is able to generate the optimal wavefront that would focus on the target through the diffuser, is to spatially decompose the beam into an array of localized laser modes (Supplementary Fig.2c). The process of lasing through the small pinhole target is then simply the process of frequency and phase locking of an array of spatially separated laser modes, a common and well established feature of coupled lasers[1-3].

In the degenerate (4-f) cavity design (Fig.1a), in the absence of a pinhole (and finite lenses and mirrors), after each cavity roundtrip, each localized laser mode is re-imaged onto itself and remains localized, as it is an eigenmode of the degenerate cavity[4]. By adding the small pinhole target in the focal plane, these modes are no longer localized as they are diffracted by propagating through the pinhole. This in turn couples these different localized modes through the process of mutual light injection. In this manner the laser array lasing solution "super mode", which focuses through the pinhole, is the solution which possess minimum loss and thus dominates all other modes through the process of mode competition.

To simulate this numerically, we spanned the laser-modes in a basis of small round modes arranged on a triangular lattice, which enables tight packing. We selected the size of these modes to be 40μm which is larger than the diffraction-limited size for our experimental system (10μm) given by the gain rod diameter (10*mm*) and length (11*cm*). We then calculated, for each specific lasing frequency within the entire free spectral range of the cavity, the steady state solution of the entire array of laser modes, taking into account the effects of coupling between modes as well as the effects of gain saturation, which allow mode competition to take place. Finally, the lasing pattern, as measured experimentally by a camera with an integration time much longer than the lasing bandwidth, is calculated by summing the intensity patterns of all the lasing modes at the different lasing frequencies. To find the steady-state electric field distributions of each laser mode, a Fox-Li type algorithm[5] was used to iteratively propagate an initial state (a random electric field distribution) again and again through the cavity; mimicking the effect of the circulating light, until eventually a steady state solution is achieved. This particular state corresponds to the largest eigenmode of the cavity super-mode.

The iterative numerical propagation in the cavity takes into account the effects of mode-competition via gain saturation, the modulation of the effective cavity length by the diffuser, and mode coupling by the small target size. The propagation process is represented schematically in Supplementary Figure 1.

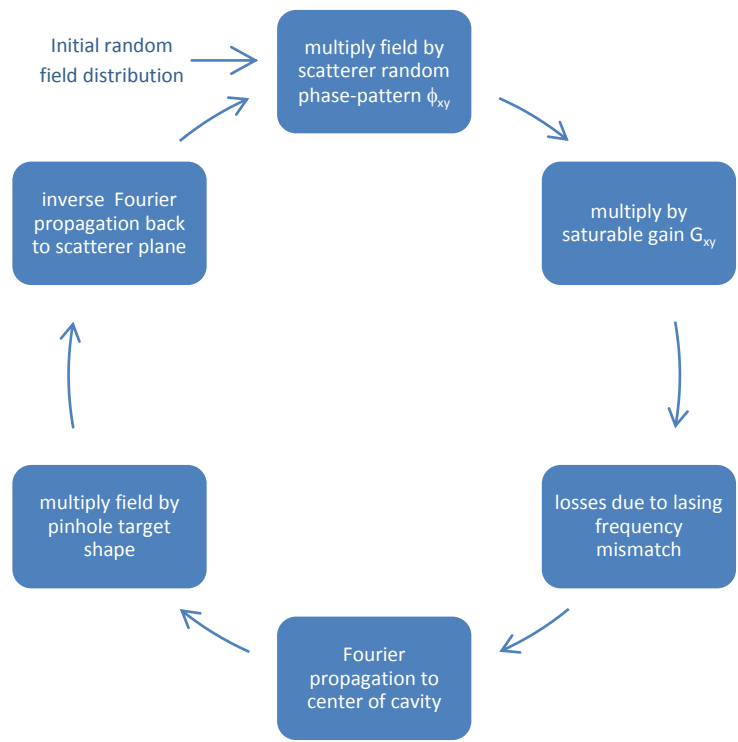

**Supplementary Figure 1.** Schematic description of the numerical field propagation steps in one cavity roundtrip, used to determine the lasing steady-state

The calculation was implemented in Matlab by Fourier propagating a two dimensional matrix $E_{ij}$ representing the spatial distribution of the field from one cavity end where it was multiplied index-wise by two dimensional matrices that represent the gain aperture and the random phase mask of the diffuser, to the cavity center where it was multiplied by the pinhole target aperture. In these steps we took into account the gain saturation by multiplying by a saturable gain matrix, $G_{ij}$,

$$G_{ij} = p / (1 + I_{ij} / I_{Sat})  \qquad (1$$

where $p$ is a constant giving the pump parameter, $I_{ij}=|E_{ij}|^2$ is the intensity distribution matrix, and $I_{Sat}$ the saturation intensity of the gain medium. We then used a two dimensional fast Fourier transform algorithm to propagate the field from the random phase mask plane to the focal plane. To simulate the effect of the different longitudinal modes we calculated, for each laser mode individually its effective optical cavity length that results from the random phase mask. This was determined by calculating the mean phase of each mode overlap with itself after one cavity roundtrip $\phi_{ij}^{diff}$. Consequently, the frequency that each localized laser mode will operate in when it is uncoupled from any other mode (e.g. in the absence of a pinhole), is given by $\omega_{ij} = 2\pi n - \phi_{ij}^{diff} / c2L$ where $c$ is the speed of light and $L$ is the cavity length. In the actual experiment, the pinhole target couples the modes by diffracting light from one mode to another; thereby the modes may lock onto the same common frequency through the process of frequency pulling. This will occur when the coupling strength $\kappa$, which is the amount of light injected from one laser to another, is sufficient to overcome the frequency detuning $\Delta\omega^{2, 6}$). Intuitively, the process is a balance between the laser on one hand, gaining energy by locking two modes to a coherent phase at a detuned frequency that coincided with the injected coupling signal, and on the other hand, suffering a roundtrip loss due to this detuning, which is proportional to[6],

$$1-\cos(\Delta\omega \cdot c2L) \qquad (2$$

To simulate the effect of simultaneous lasing at several different frequencies at the longitudinal modes available for each localized laser mode, we simply took into account the roundtrip losses associated with frequency mismatch in accordance with Eq.(2). Specifically, for a given lasing frequency $\omega_0$ we multiply the array by the loss matrix:

$$L_{ij} = 1 - \cos(\phi_{ij}^{diff} + \phi_0). \qquad (3$$

where $\phi_0 = \omega_0 c 2L$ denotes the phase accumulated by propagating back and forth through one roundtrip of the cavity.

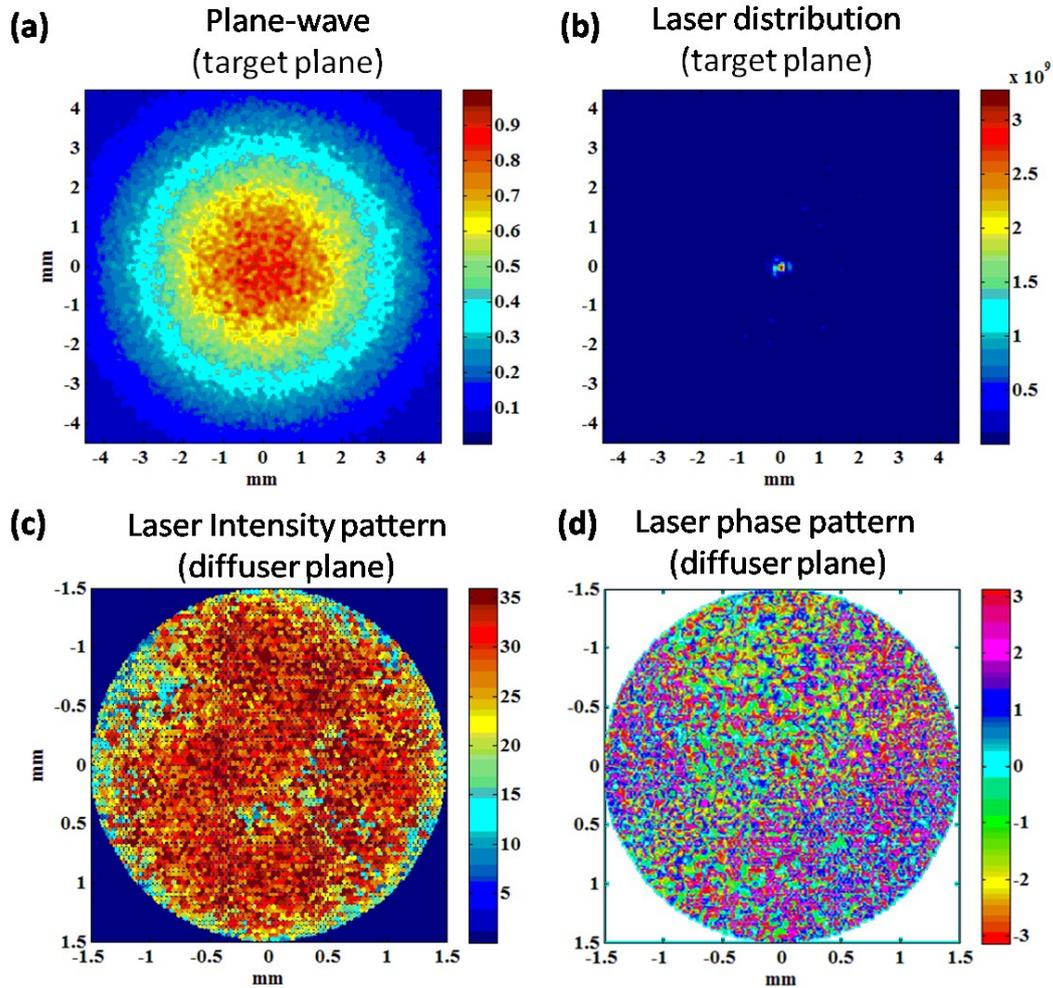

**Supplementary Figure 2. Simulation result of focusing light through a randomly scattering media. (a)** and **(b)** Show the simulated intensity distributions of the focused light that passed through a random phase mask with similar characteristics to the experimental diffuser. **(a)** Shows the simulation results with incident plane wave illumination, **(b)** shows the simulated laser pattern. **(c)** and **(d)** Show the simulated intensity and corresponding phase distributions of the laser mode which illuminates the diffuser.

The numerical results obtained with this model using the experimental parameters are presented in Supplementary Fig. 2. To account for the finite pumping region of our flash-lamp (in the experimental apparatus) we used a 3 mm diameter gain that consisted of over 6000 modes. Supplementary Figs. 2(a) and 2(b) show the intensity distribution at the target plane. For a reference plane-wave (Supplementary Fig. 2(a)) scattered by the random diffuser phase mask the speckle pattern has a typical spread that resembles the experimental results in Fig. 2(a) of the manuscript. In contrast, for the simulated laser "super mode" intensity pattern that is passing as well through the same random phase mask, tight focusing that closely resembles the experimental result occurs (Supplementary Fig. 2(b)). Moreover, a quantitative comparison of the intensity cross-sections for the experimental and simulated lasing patterns shows good match. Supplementary Figs. 2(c) and 2(d) show the intensity (Supplementary Fig. 2(c)) and phase (Supplementary Fig. 2(d)) distributions at the diffuser plane for the same lasing "super mode" of Supplementary Fig. 2(b). As evident, the laser exploits the shaping of both the amplitudes as well as the phases of a large portion of the lasers in the array in order to focus, which gives it the large number of degrees of freedom that is required to generate the complex wavefront pattern and the high-contrast focus.

## References

1. Nixon, M. et al. Synchronized cluster formation in coupled laser networks. *Physical Review Letters* **106**, 223901-223901 (2011).
2. Fabiny, L., Colet, P., Roy, R. & Lenstra, D. Coherence and phase dynamics of spatially coupled solid-state lasers. *Physical Review A;(United States)* **47** (1993).
3. Fan, T. Laser beam combining for high-power, high-radiance sources. *Selected Topics in Quantum Electronics, IEEE Journal of* **11**, 567-577 (2005).
4. Arnaud, J. Degenerate optical cavities. *Applied optics* **8**, 189-195 (1969).
5. Fox, A. & Li, T. Modes in a maser interferometer with curved and tilted mirrors. *Proceedings of the IEEE* **51**, 80-89 (1963).
6. Siegman, A.E. Lasers, Edn. 1. ( University Science Books, Mill Valley, CA 94941; 1986).